\documentclass[aps,twocolumn]{revtex4-1}
\usepackage{hyperref}
\usepackage[sort&compress]{natbib}
\usepackage{amsfonts}
\usepackage{amsmath}
\usepackage{amssymb,epsf}
\usepackage{latexsym}
\usepackage{graphicx,epsfig}
\usepackage{epstopdf}
\usepackage{caption}
\usepackage{color}                                    

\begin{document}

\title{Kaniadakis Holographic Dark Energy with Particle Horizon as IR Cutoff}
\author{A. Asvar$^{1}$\footnote{a_asvar@pnu.ac.ir}, M. Mohammadi$^{2}$\footnote{physmohammadi@pgu.ac.ir},
A. Sheykhi$^{1}$\footnote{asheykhi@shirazu.ac.ir}}
\affiliation{$^1$ Department of Physics, College of Science, Shiraz University, Shiraz 71454, Iran\\
$^2$ Department of Physics, Persian Gulf University, Bushehr 75169, Iran}

\begin{abstract}
We construct a holographic dark-energy model using Kaniadakis
entropy with the particle horizon as the infrared cutoff,
consistently modifying both the holographic density and the
Friedmann background. In standard Einstein gravity, noninteracting
particle-horizon holographic dark energy (HDE) does not produce the
sufficiently negative pressure required for late-time accelerated
expansion. We show that the Kaniadakis deformation changes this
behavior while the particle horizon is retained. For the
representative parameter choice $K=1.90\times10^{-36}$, with
$\Omega_{\rm DE0}=0.7$ and $c^2=0.64$, the deceleration parameter
changes sign at $z\simeq0.585$ in the noninteracting case. Including
the interaction strengths $b^2=0.03$ and $0.06$ changes the
transition only slightly, giving $z\simeq0.586$ and $0.589$,
respectively. Thus, for the parameter set considered here, the
interaction does not generate the acceleration; rather, the
transition is already present in the noninteracting Kaniadakis
model and the interaction produces only a small shift in its timing.
We derive the autonomous evolution equations, the effective
dark-energy equation of state, and the deceleration parameter.
The adiabatic squared sound speed is also examined as a diagnostic
of the effective-fluid stability, while statefinder variables are
used to characterize deviations from $\Lambda$CDM. Finally, we
verify that the $K\to0$ limit continuously recovers standard
Einstein-gravity particle-horizon HDE, for which the noninteracting
model remains decelerating.

\end{abstract}

\maketitle

\section{Introduction}\label{sec:intro}
The discovery of the late-time accelerated expansion of the
Universe through observations of Type-Ia supernovae
\cite{Riess1998,Perlmutter1999} established one of the central
problems of modern cosmology: identifying the physical origin of
the component responsible for cosmic acceleration. The
cosmological constant provides the simplest and observationally
successful explanation within the $\Lambda$CDM framework.
Nevertheless, the extremely small observed value of the
vacuum-energy scale and the coincidence between the present matter
and dark-energy densities continue to motivate the investigation
of dynamical alternatives
\cite{Weinberg1989,Carroll2001,Padmanabhan2003}. Among these
alternatives, HDE is particularly appealing because it relates the
cosmological dark-energy scale to fundamental bounds on the number
of gravitational degrees of freedom.

The conceptual origin of HDE lies in the holographic principle.
Bekenstein demonstrated that the entropy of a gravitating system
is fundamentally constrained by the area of its boundary and that
the entropy of a black hole is proportional to its horizon area
\cite{Bekenstein1973}. This insight was subsequently developed
into the holographic principle by 't Hooft and Susskind, who
argued that the fundamental number of degrees of freedom in a
gravitational system scales with the area rather than the volume
of the region \cite{tHooft1993,Susskind1995}. Covariant
formulations of holographic entropy bounds were later developed by
Bousso \cite{Bousso2002}. These developments provided the
conceptual framework for applying gravitational entropy bounds to
cosmology.

A direct connection between holography and dark energy was
established by Cohen, Kaplan, and Nelson \cite{Cohen}. Their
argument is based on the observation that the ultraviolet cutoff
of an effective field theory cannot be chosen independently of its
infrared size: the total energy contained in a region of
characteristic length $L$ should not be sufficient to produce a
black hole of that size. This UV--IR relation leads, at the level
of scaling, to an upper bound on the vacuum-energy density that
decreases with the square of the infrared length. Interpreting the
largest allowed energy density as a cosmological dark-energy
component leads to the standard holographic dark-energy
prescription. Thus, the origin of HDE is not an arbitrary
modification of the cosmological energy budget, but rather the
combination of the holographic principle, gravitational entropy,
and the requirement that an effective field theory remain below
the black-hole formation bound.

The subsequent development of HDE demonstrated that the choice of
the infrared cutoff is crucial for its cosmological behavior. Hsu
showed that identifying the infrared scale with the Hubble radius
does not provide a sufficiently negative equation of state in the
simplest HDE construction \cite{Hsu2004}. Li subsequently proposed
the future event horizon as the infrared cutoff and obtained a
phenomenologically viable accelerating model \cite{Li2004}. This
result stimulated extensive investigations of alternative infrared
prescriptions, spatial curvature, modified gravitational dynamics,
and interactions between dark energy and dark matter
\cite{HuangGong2004,Pavon2005,Gao2009,Granda2009,Wang2017}. The
freedom in choosing the infrared cutoff is therefore one of the
defining features of HDE phenomenology.

Interacting HDE models have received particular attention because
an energy transfer between the dark sectors can substantially
modify the cosmic expansion history. Wang, Gong, and Abdalla
studied the transition of the dark-energy equation of state in an
interacting HDE framework \cite{WangGong2005}, while Wang, Lin,
and Abdalla investigated observational constraints on such models
\cite{WangLin2006}. A broad review of dark matter--dark energy
interactions and their theoretical and observational consequences
was subsequently presented by Wang \textit{et al.}
\cite{WangReview2016}. These studies established that dark-sector
interaction can significantly alter the effective equation of
state and the timing of the deceleration-to-acceleration
transition. At the same time, an important conceptual distinction
exists between acceleration produced by an explicit energy
exchange between dark matter and dark energy and acceleration
arising from a modification of the holographic or gravitational
sector itself. This distinction is particularly relevant when
assessing generalized entropy corrections.

Other holographic-inspired dark-energy models have explored
characteristic cosmological time scales as alternatives to
horizon-based infrared cutoffs. Agegraphic dark energy, for
example, relates the dark-energy scale to the age of the Universe
\cite{Cai2007}, while new agegraphic dark energy uses the
conformal time as the relevant cosmological scale
\cite{WeiCai2008}. These developments illustrate the broader
possibility that the infrared scale entering an effective
dark-energy density can be associated with different geometrical
or temporal properties of the cosmological spacetime. In the
present work, we retain the particle horizon as the infrared
cutoff and instead modify the entropy and gravitational sectors.

The particle horizon is a particularly interesting choice because
it is naturally constructed from the past light cone. Its
cosmological behavior differs fundamentally from that of the
future event horizon. In standard HDE based on the usual
Bekenstein--Hawking entropy and Einstein gravity, however, the
particle horizon does not provide a sufficiently negative
dark-energy equation of state in the noninteracting case
\cite{Li2004,Wang2017}. Consequently, standard noninteracting
particle-horizon HDE does not produce the required late-time
accelerated expansion. This result is important because it
establishes a clear reference point for the present investigation.
If accelerated expansion can be obtained while the particle
horizon is retained, the origin of the new behavior must lie
elsewhere, for example in the entropy or gravitational sector,
rather than in a change of the infrared cutoff.

A natural way to investigate this possibility is to modify the
entropy entering the holographic construction. Several generalized
entropy prescriptions have been incorporated into HDE, including
logarithmic and power-law corrections to the Bekenstein--Hawking
entropy \cite{Wei2009,Sheykhi2011}, as well as Tsallis, R\'{e}nyi,
and Barrow entropy-based constructions
\cite{TsallisCirto2013,Tavayef2018,Zadeh2018,Moradpour2018,Barrow2020,Saridakis2020,Anagnostopoulos2020}.
These approaches share a common idea: the holographic principle is
retained, while the entropy-area relation underlying the
gravitational bound is generalized.

Kaniadakis statistics provides a particularly interesting
deformation of the standard statistical framework. Kaniadakis
introduced a one-parameter relativistic generalization of
Boltzmann--Gibbs statistics, motivated by the requirements of
relativistic composition laws and a consistent nonextensive
statistical description \cite{Kaniadakis2002,Kaniadakis2005}. An
important feature of this deformation is its continuous connection
to the standard statistical theory: the usual entropy is recovered
when the deformation parameter vanishes. When the Kaniadakis
framework is applied to horizon thermodynamics, the corresponding
black-hole entropy acquires nonlinear corrections to the standard
Bekenstein--Hawking expression. These corrections provide a
controlled way of investigating how departures from the
conventional entropy-area relation can influence cosmological
dynamics.

The cosmological implications of Kaniadakis entropy have been
investigated in both modified-gravity and holographic settings.
Lymperis, Basilakos, and Saridakis derived modified cosmological
equations associated with Kaniadakis horizon entropy
\cite{Lymperis2021}. Kaniadakis HDE was subsequently investigated
using different entropy and infrared prescriptions
\cite{Moradpour2020,Drepanou2022}. More recently, explicit
Kaniadakis corrections to the Friedmann equations have been
derived, showing that the entropy deformation can modify not only
the holographic energy density but also the gravitational
background itself \cite{Sheykhi2024}. Related studies have also
examined HDE in modified Kaniadakis cosmology with the Hubble
radius as the infrared cutoff, including its background evolution,
dark-sector interaction, stability, and statefinder diagnostics
\cite{Sheykhi2026}.

This last point is essential for the construction considered here.
If the Kaniadakis correction is introduced only into the
holographic energy density while the standard Einstein Friedmann
equation is retained, only part of the physical effect associated
with the entropy deformation is taken into account. In a
thermodynamically consistent Kaniadakis framework, the modified
horizon entropy also affects the gravitational background
equations \cite{Lymperis2021,Sheykhi2024}. We therefore
incorporate the Kaniadakis correction consistently in both
sectors: the holographic dark-energy density and the Friedmann
background. This procedure produces an effective dark-energy
sector containing the conventional inverse-square holographic
contribution, a Kaniadakis correction with a different dependence
on the particle-horizon scale, and an additional geometric
contribution originating from the modified gravitational dynamics.
The explicit expressions are derived in Sec.~\ref{sec:model}.

The central question of this paper is therefore more specific than
the general question of whether Kaniadakis entropy can modify HDE.
We ask whether the Kaniadakis deformation can change the
well-known nonaccelerating behavior of \emph{noninteracting
particle-horizon HDE in standard Einstein gravity}, while
retaining the particle horizon itself as the infrared cutoff. This
provides a particularly clean test of the physical role of the
entropy deformation because the infrared prescription is kept
unchanged and no energy transfer between dark matter and dark
energy is initially introduced. We find that the answer is
affirmative for the representative parameter set considered in this
work. We fix the Kaniadakis deformation parameter to
$K=1.90\times10^{-36}$ for the numerical analysis; this value is
adopted as a representative choice for which the deceleration-to-
acceleration transition occurs close to $z\simeq0.6$. For
$\Omega_{\rm DE0}=0.7$ and $c^2=0.64$, the noninteracting model
has $q=0$ at $z\simeq0.585$. The interacting cases with
$b^2=0.03$ and $0.06$ give $z\simeq0.586$ and $0.589$,
respectively. Hence, the finite-$K$ deformation is responsible for
allowing acceleration in the particle-horizon model, while the
interaction considered here produces only a very small change in
the transition redshift. This is the main physical result of the
present work.

We further investigate the physical behavior of the solutions
through the adiabatic squared sound speed and the statefinder
diagnostics. The sound-speed analysis provides a useful diagnostic
of the stability of the effective dark-energy fluid, while the
statefinder variables provide a geometrical characterization of
departures from $\Lambda$CDM \cite{Sahni2003,Alam2003,Zhang2006}.
The combined analysis allows us to distinguish the effects of the
Kaniadakis deformation from those of the dark-sector interaction
and to determine whether the parameter choices that produce
accelerated expansion are accompanied by potentially problematic
stability behavior.

The paper is organized as follows. In Sec.~\ref{sec:model}, we
introduce the Kaniadakis entropy deformation, construct the
corresponding particle-horizon holographic dark-energy density,
and derive the modified Friedmann equations. In
section~\ref{sec:dynamics}, we develop the autonomous cosmological
dynamics, including the physical particle-horizon branch and the
interacting extension. In section~\ref{sec:results}, we present
the numerical background evolution, stability analysis, and
statefinder diagnostics. In Sec.~\ref{sec:limits}, we discuss the
undeformed and noninteracting limits and clarify the connection
with standard Einstein-gravity particle-horizon HDE. Finally, In
section ~\ref{sec:conclusion}, we summarize the main results and
discusses their physical significance and possible extensions.
\section{Kaniadakis holographic dark energy}\label{sec:model}
\subsection{Entropy deformation and holographic density}
The Kaniadakis entropy can be expressed as
\cite{Kaniadakis2002,Kaniadakis2005}
\begin{equation}\label{eq:SK}
S_K=-\sum_i\frac{P_i^{1+K}-P_i^{1-K}}{2K},
\end{equation}
where $K$ is the dimensionless deformation parameter. The standard
Boltzmann-Gibbs form is recovered when $K\to0$. For the black-hole
application, taking the microstate probabilities to be equal and
using the Bekenstein--Hawking entropy $S_{\rm BH}=A/(4G)$ gives
\begin{equation}\label{eq:SKBH}
S_K=\frac{1}{K}\sinh(KS_{\rm BH}) \simeq S_{\rm
BH}+\frac{K^2}{6}S_{\rm BH}^3+\mathcal{O}(K^4),
\end{equation}
where in the last step we have expanded $S_K$ by assuming that
$K\ll1$. Thus the leading correction is even in $K$. Consequently,
all background quantities obtained from the leading holographic
correction depend on $K^2$, and the analysis may be restricted to
$K\geq0$ without loss of generality.

The holographic bound may be written schematically as $\rho_{\rm
DE}L^4\lesssim S$. Substitution of the Kaniadakis entropy and
retention of the leading nonstandard contribution yields
\cite{Drepanou2022}
\begin{equation}\label{eq:rhoKHDE}
\rho_{\rm DE}^{(H)}=3c^2M_p^2L^{-2}+K^2M_p^6L^2,
\end{equation}
where $c$ is the usual dimensionless HDE parameter and an
inessential numerical factor has been absorbed into the definition
of $K$. The first term is precisely the standard HDE contribution,
while the second is the leading Kaniadakis correction. Hence the
standard HDE density is recovered continuously for $K\to0$.

For the infrared scale we choose the particle horizon,
\begin{equation}\label{eq:RP}
R_P=a\int_0^t\frac{dt'}{a(t')}
=a\int_0^a\frac{da'}{Ha'^2},
\end{equation}
which obeys
\begin{equation}\label{eq:RPdot}
\dot R_P=HR_P+1.
\end{equation}
\subsection{Kaniadakis-modified Friedmann equations}
The modification of the horizon entropy also changes the
gravitational background equations. For a spatially flat universe
the Kaniadakis-corrected Friedmann equations can be written as
\cite{Sheykhi2024}
\begin{align}
3M_p^2\left(H^2-\alpha H^{-2}\right)&=\rho_m+\rho_{\rm DE}^{(H)},\label{eq:F1}\\
-2M_p^2\dot H\left(1+\alpha H^{-4}\right)
&=\rho_m+p_m+\rho_{\rm DE}^{(H)}+p_{\rm DE}^{(H)},\label{eq:F2}
\end{align}
where
\begin{equation}\label{eq:alpha}
\alpha=\frac{K^2\pi^2}{2G^2}.
\end{equation}
The same deformation parameter therefore enters both the
holographic energy density and the gravitational background
dynamics. Treating only the first effect while retaining standard
Friedmann dynamics would not describe the full
Kaniadakis-corrected background. For the cosmological analysis it
is convenient to move the geometric correction in
Eq.~(\ref{eq:F1}) to the energy-density side. We therefore define
\begin{equation}\label{eq:rhoeff}
\rho_{\rm DE}\equiv\rho_{\rm DE}^{\rm eff}
=3c^2M_p^2R_P^{-2}+K^2M_p^6R_P^2
+3\alpha M_p^2H^{-2}.
\end{equation}
The Friedmann constraint then takes the standard-looking form
\begin{equation}\label{eq:Fstandard}
3M_p^2H^2=\rho_m+\rho_{\rm DE}.
\end{equation}
Defining
\begin{equation}\label{eq:Omega}
\Omega_i=\frac{\rho_i}{3M_p^2H^2},
\end{equation}
we obtain
\begin{equation}\label{eq:closure}
\Omega_m+\Omega_{\rm DE}=1.
\end{equation}
This rewriting does not introduce an additional physical fluid.
Indeed, it is a bookkeeping device that places the geometric
Kaniadakis correction into the effective dark-energy sector.
\section{Cosmological dynamics}\label{sec:dynamics}
We use $x=\ln a$ as the independent variable, so that a prime
denotes $d/dx$. For pressureless matter, $p_m=0$ and
\begin{equation}\label{eq:rhom}
\rho_m=\rho_{m0}a^{-3},\qquad
\Omega_m=1-\Omega_{\rm DE}.
\end{equation}
Consequently,
\begin{equation}\label{eq:Ha}
\frac{1}{Ha}=\frac{\sqrt{a(1-\Omega_{\rm DE})}}
{H_0\sqrt{\Omega_{m0}}}.
\end{equation}
\subsection{Particle-horizon branch and autonomous equation}
Substituting Eq.~(\ref{eq:rhoeff}) into the modified Friedmann
constraint gives a quadratic equation in $R_P^2$,
\begin{equation}\label{eq:RPpoly}
K^2M_p^6R_P^4-AR_P^2+3c^2M_p^2=0,
\end{equation}
where
\begin{equation}\label{eq:A}
A=3M_p^2(H^2-\alpha H^{-2})-\rho_m.
\end{equation}
The physical branch is selected as
\begin{equation}\label{eq:RPbranch}
R_P^2=\frac{A-\sqrt{A^2-12K^2c^2M_p^8}}
{2K^2M_p^6}.
\end{equation}
Expanding the square root for small $K$ shows that the numerator
vanishes as $K^2$, leaving a finite result. The opposite algebraic
root diverges as $K\to0$ and therefore has no continuous
standard-HDE limit. The branch choice is consequently fixed by the
physical requirement of continuity with standard particle-horizon
HDE. For numerical work it is convenient to introduce
\begin{eqnarray}
A_D&=&\frac{A+\sqrt{A^2+36\alpha\Omega_{\rm DE}}}{2},\nonumber\\
\qquad
B_D&=&\frac{A-\sqrt{A^2-12K^2c^2M_p^4}} {2K^2M_p^4},\\
\alpha_D&=&\frac{9\alpha\Omega_{\rm DE}}{A_D^2}.
\end{eqnarray}
With these definitions, differentiating the particle-horizon
relation and using the matter evolution yields
\begin{equation}\label{eq:OmegaCompact}
\Omega_{\rm DE}'=
\frac{\Omega_{\rm DE}(1-\Omega_{\rm DE})}
{1+\alpha_D\Omega_{\rm DE}}\,\mathcal F_0,
\end{equation}
where
\begin{equation}\label{eq:F0}
\mathcal F_0\equiv3(1+\alpha_D) -\frac{2(A-2K^2M_p^4B_D)}{A_D}
\left[1+\sqrt{\frac{3\Omega_{\rm DE}}{A_DB_D}}\right].
\end{equation}
The redshift equation is
\begin{equation}\label{eq:Omegaz}
\frac{d\Omega_{\rm DE}}{dz}=-\frac{\Omega_{\rm DE}'}{1+z}.
\end{equation}
The dark-energy equation of state follows from the continuity
equation. For the noninteracting case,
\begin{equation}\label{eq:continuityDE}
\dot\rho_{\rm DE}+3H(1+w_{\rm DE})\rho_{\rm DE}=0,
\end{equation}
so that
\begin{equation}\label{eq:wDEcompact}
w_{\rm DE}=-1-\frac{\dot\rho_{\rm DE}}
{3H\rho_{\rm DE}}.
\end{equation}
Defining
\begin{equation}\label{eq:LambdaH}
\Gamma_H\equiv-6\alpha H^{-3}\dot H =-3\alpha H^{-1}
\left(-3+\frac{\Omega_{\rm DE}'}{1-\Omega_{\rm DE}}\right),
\end{equation}
and using $H=\sqrt{A_D/(3\Omega_{\rm DE})}$, one obtains
\begin{equation}\label{eq:wDEexplicit}
\begin{aligned}
w_{\rm DE}={}&-1
-2\left(\frac{\Omega_{\rm DE}}{3A_D^3}\right)^{1/2}
\left(\frac{-3c^2+K^2M_p^4B_D^2}{B_D^{3/2}}\right)\\
&\times\left[1+\frac{\sqrt{3}}{3} \left(\frac{A_DB_D}{\Omega_{\rm
DE}}\right)^{1/2}\right] -\Gamma_H\left(\frac{\Omega_{\rm
DE}}{3A_D^3}\right)^{1/2}.
\end{aligned}
\end{equation}
In the standard limit $K\to0$, the Kaniadakis corrections vanish
and the particle-horizon HDE result is recovered. In particular,
in the dark-energy-dominated limit,
\begin{equation}
\left.w_{\rm DE}\right|_{K\to0}=-\frac13+\frac{2}{3c}>-\frac13,
\end{equation}
which explicitly demonstrates why standard noninteracting
particle-horizon HDE does not accelerate in Einstein gravity
\cite{Li2004,Wang2017}. The derivative of Eq.~(\ref{eq:rhoeff}) is
evaluated using relation (\ref{eq:RPdot}). We find
\begin{equation}\label{eq:rhodot}
\dot\rho_{\rm DE}=2M_p^2
\left(-3c^2R_P^{-3}+K^2M_p^4R_P\right)\dot R_P
-6\alpha M_p^2H^{-3}\dot H.
\end{equation}
The total equation of state and deceleration parameter are
\begin{equation}\label{eq:wtot}
w_{\rm tot}=w_{\rm DE}\Omega_{\rm DE},
\qquad
q=\frac12+\frac32w_{\rm DE}\Omega_{\rm DE}.
\end{equation}
Acceleration starts when $q$ changes sign, equivalently when
$w_{\rm tot}=-1/3$. A useful limiting check is obtained by setting
$K=0$. In that case the Kaniadakis terms disappear and the
particle-horizon model reduces to standard HDE in standard
Einstein gravity. Since its noninteracting equation of state is
not sufficiently negative, the standard model is decelerating. The
accelerated solution found below is therefore not a consequence of
the particle horizon alone: it is a consequence of the Kaniadakis
entropy deformation while the same particle horizon is retained.
\subsection{Interaction between the dark sectors}
We next allow energy exchange between dark energy and pressureless
dark matter. We adopt
\begin{align}
\dot\rho_{\rm DE}+3H(1+w_{\rm DE})\rho_{\rm DE}&=-Q,\label{eq:contint1}\\
\dot\rho_m+3H\rho_m&=Q,\label{eq:contint2}
\end{align}
with
\begin{equation}\label{eq:Q}
Q=3b^2H(1+r)\rho_{\rm DE},
\qquad
r=\frac{1-\Omega_{\rm DE}}{\Omega_{\rm DE}}.
\end{equation}
Positive $b^2$ corresponds to transfer of energy from dark energy
to dark matter under this sign convention. The total conservation
law remains unchanged.  The autonomous equation becomes
\begin{equation}\label{eq:OmegaInt}
\Omega_{\rm DE}'=
\frac{\Omega_{\rm DE}(1-\Omega_{\rm DE})}
{1+\alpha_D\Omega_{\rm DE}}
\left[\mathcal F_0+
\frac{3b^2}{\Omega_{\rm DE}}
-\frac{3b^2}{\Omega_{\rm DE}(1-\Omega_{\rm DE})}\right].
\end{equation}
The interacting equation of state is
\begin{equation}\label{eq:wDEint}
w_{\rm DE}^{\rm int}=w_{\rm DE}-\frac{b^2}{\Omega_{\rm DE}},
\end{equation}
where $w_{\rm DE}$ is the noninteracting result. Thus the
interaction shifts the effective equation of state toward more
negative values for positive $b^2$. The deceleration parameter is
\begin{equation}\label{eq:qint}
q=\frac12+\frac32w_{\rm DE}^{\rm int}\Omega_{\rm DE}.
\end{equation}
When $b^2=0$, the noninteracting Kaniadakis particle-horizon
equations are recovered. When both $b^2=0$ and $K=0$, standard
particle-horizon HDE in Einstein gravity is recovered, and the
nonaccelerating result follows.
\section{Results and diagnostics}\label{sec:results}
We numerically integrate Eqs.~(\ref{eq:OmegaCompact}) and
(\ref{eq:OmegaInt}) from the present epoch toward the past, using
$\Omega_{\rm DE0}=0.7$ and $c^2=0.64$. The interaction strengths
considered are $b^2=0$, $0.03$, and $0.06$. Throughout the numerical
analysis we fix the representative Kaniadakis parameter to
$K=1.90\times10^{-36}$. This value is adopted as a representative
choice that places the resulting deceleration-to-acceleration
transition close to $z\simeq0.6$; once fixed, the same value is used
consistently in all figures and diagnostics.

\begin{figure}[t]
\centering
\includegraphics[width=0.32\textwidth]{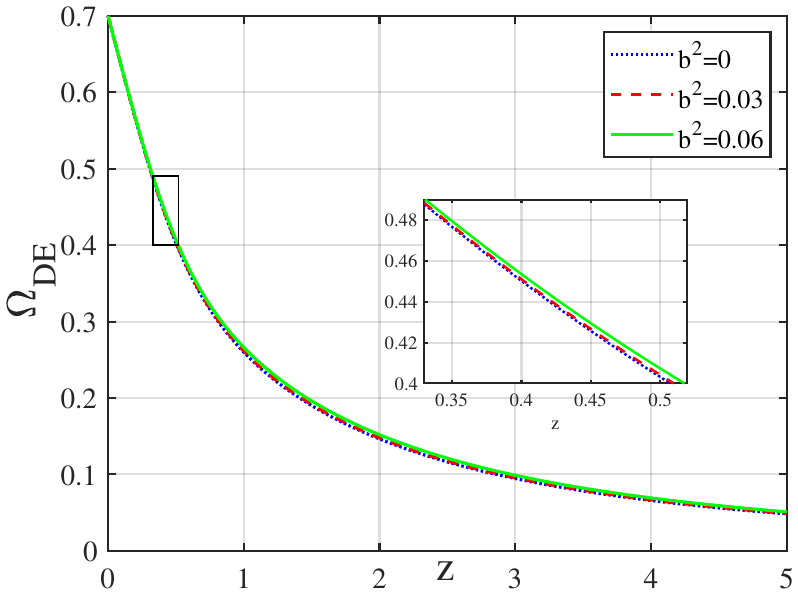}
\caption{Evolution of the dark-energy density parameter $\Omega_{\rm DE}(z)$. The inset magnifies the low-redshift evolution over a restricted interval.}
\label{fig:OmDE}
\end{figure}

\begin{figure}[t]
\centering
\includegraphics[width=0.32\textwidth]{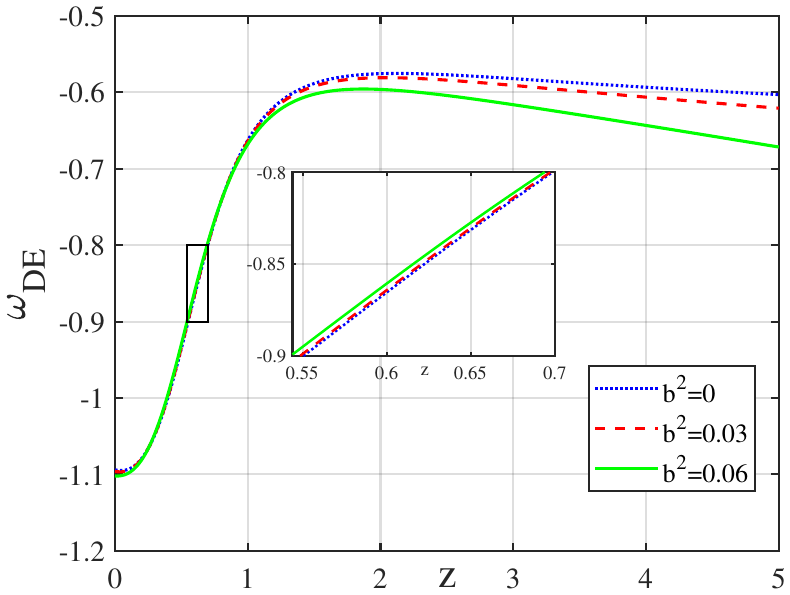}
\caption{Evolution of the dark-energy equation-of-state parameter \(w_{\rm DE}(z)\). The inset provides an enlarged view of the narrow redshift interval \(0.55\lesssim z\lesssim0.70\).}
\label{fig:wDE}
\end{figure}

\begin{figure}[t]
\centering
\includegraphics[width=0.32\textwidth]{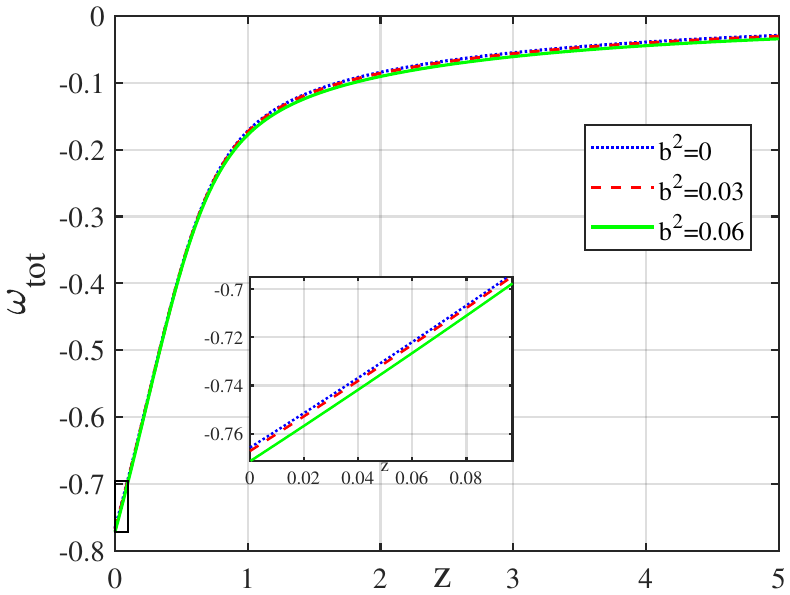}
\caption{Evolution of the total equation-of-state parameter $w_{\rm tot}(z)$. The inset magnifies the interval $0\le z\lesssim0.10$.}
\label{fig:wtot}
\end{figure}

\begin{figure}[t]
\centering
\includegraphics[width=0.32\textwidth]{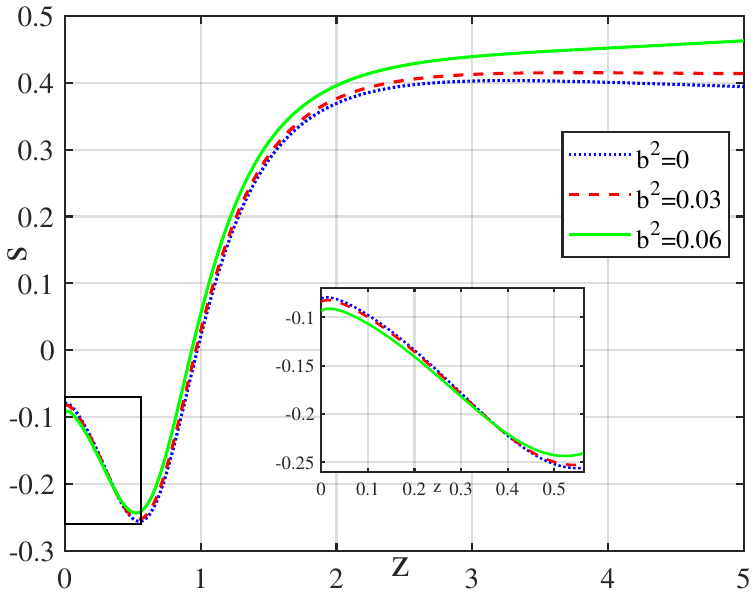}
\caption{Statefinder parameter $s(z)$. The inset magnifies the interval $0\le z\lesssim0.55$.} \label{fig:s}
\end{figure}

\begin{figure}[t]
\centering
\includegraphics[width=0.32\textwidth]{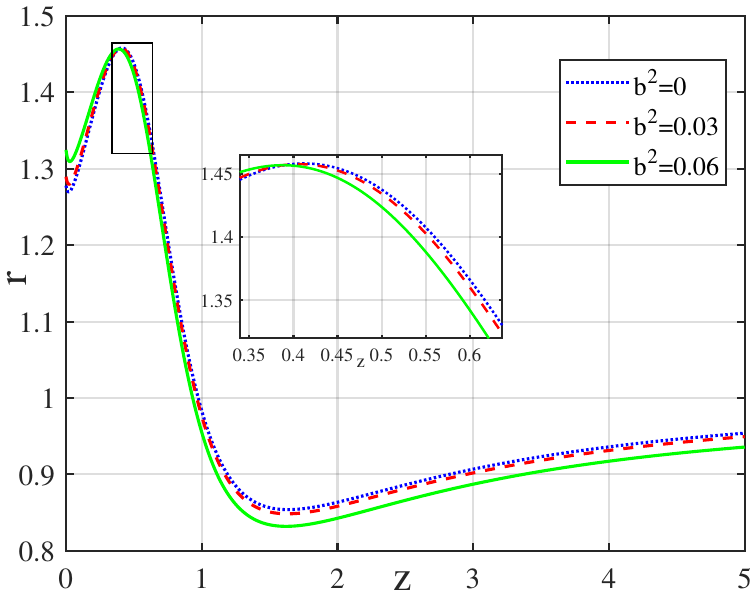}
\caption{Statefinder parameter $r(z)$. The inset magnifies the low-redshift maximum around $z\simeq0.4$--$0.65$.} \label{fig:r}
\end{figure}

The solutions are selected on the physical branch of
Eq.~(\ref{eq:RPbranch}) throughout the integration domain. The
figures display the resulting evolution for $0\le z\le5$. The
parameter set stated above, namely $\Omega_{\rm DE0}=0.7$,
$c^2=0.64$, and $K=1.90\times10^{-36}$, is used throughout
Figs.~\ref{fig:OmDE}--\ref{fig:sr}. The three curves in each
figure correspond to $b^2=0$, $0.03$, and $0.06$, as indicated in
the legends; these interaction values are not repeated in every
caption.

\subsection{Background evolution}
\begin{figure}[t]
\centering
\includegraphics[width=0.32\textwidth]{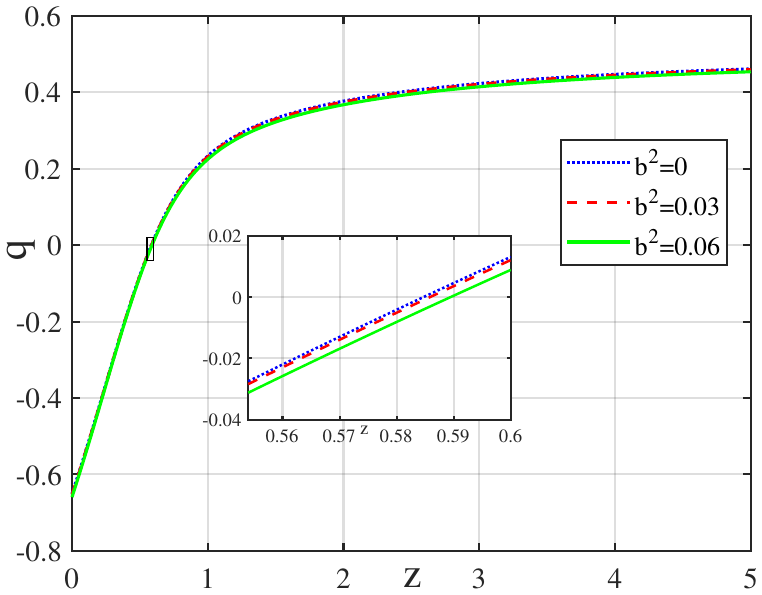}
\caption{Deceleration parameter $q(z)$. The inset magnifies the zero-crossing region and shows that $q=0$ occurs at $z\simeq0.585$, $0.586$, and $0.589$ for $b^2=0$, $0.03$, and $0.06$, respectively.}
\label{fig:q}
\end{figure}

The fractional dark-energy density decreases toward higher
redshift, so dark energy is subdominant during the earlier
matter-dominated epoch. The present normalization $\Omega_{\rm
DE0}=0.7$ is imposed by construction. The most important feature
is the behavior of the noninteracting solution.
For $b^2=0$, the Kaniadakis particle-horizon
model evolves from a decelerating matter-dominated regime to
late-time acceleration, with $q$ changing sign at
$z\simeq0.585$. This is qualitatively different from standard
Einstein-gravity particle-horizon HDE, for which the
noninteracting model does not accelerate. The comparison therefore
isolates the role of the Kaniadakis correction: the IR cutoff is
identical in the two cases, while the entropy-induced correction
changes the background dynamics sufficiently to permit
acceleration. For $b^2=0.03$ and $0.06$, the corresponding
transition redshifts are $z\simeq0.586$ and $z\simeq0.589$,
respectively. Thus, for the parameter values considered here, the
interaction modifies the transition only very weakly and shifts it
slightly toward higher redshift. The near coincidence of the three
transition redshifts is clearly visible in the inset of Fig.~\ref{fig:q}.
These interaction results should be regarded as a secondary
extension because the principal acceleration result is already
present at $b^2=0$.

\subsection{Stability}
\begin{figure}[t]
\centering
\includegraphics[width=0.32\textwidth]{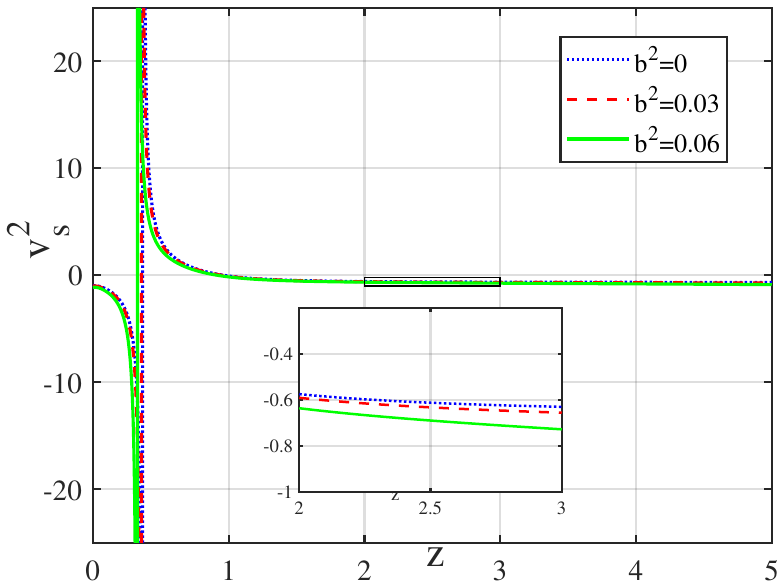}
\caption{Adiabatic squared sound speed $v_s^2(z)$. The inset magnifies the interval $2\lesssim z\lesssim3$, where the three curves remain below zero.} \label{fig:vs2}
\end{figure}

A background solution that accelerates is not by itself sufficient
to establish physical viability. We examine the adiabatic squared
sound speed,
\begin{equation}\label{eq:vs}
v_s^2=\frac{dp_{\rm DE}}{d\rho_{\rm DE}}
=w_{\rm DE}+\rho_{\rm DE}
\frac{dw_{\rm DE}/d\Omega_{\rm DE}}
{d\rho_{\rm DE}/d\Omega_{\rm DE}}.
\end{equation}
Positive $v_s^2$ corresponds to a real adiabatic propagation speed
in this effective description, while a negative value is commonly
interpreted as a classical instability of the corresponding
homogeneous-fluid description. For the present parameter choice,
the three curves in Fig.~\ref{fig:vs2} are negative over substantial
parts of the plotted redshift range, with a pronounced singular
behavior at an intermediate redshift. Increasing the interaction
strength makes the negative-$v_s^2$ behavior more pronounced,
particularly for $b^2=0.06$. Thus, although the background solution
can accelerate, the effective-fluid stability diagnostic does not
indicate complete classical stability over the full interval shown.
This result should be interpreted cautiously: $v_s^2$ is only an
adiabatic effective-fluid diagnostic and is not a substitute for a
full perturbation analysis of the underlying modified-gravity
theory.
\subsection{Statefinder diagnostic}
We use the statefinder variables \cite{Sahni2003,Alam2003}. The
first statefinder is
\begin{equation}\label{eq:r}
r=\frac{\dddot a}{aH^3}=2q^2+q-\frac{\dot q}{H},
\end{equation}
while
\begin{equation}\label{eq:s}
s=\frac{r-1}{3\left(q-\frac12\right)}.
\end{equation}
For a spatially flat $\Lambda$CDM cosmology the fixed point is $\{r,s\}=\{1,0\}$.

\begin{figure}[t]
\centering
\includegraphics[width=0.32\textwidth]{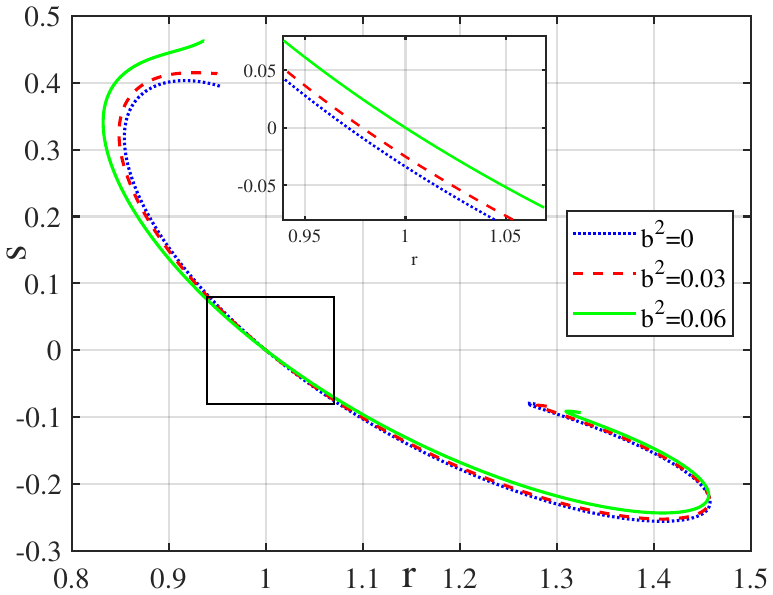}
\caption{The statefinder $s$--$r$ trajectories. The inset provides a
magnified view around the $\Lambda$CDM point $\{r,s\}=\{1,0\}$.} \label{fig:sr}
\end{figure}
For the noninteracting KHDE solution, the statefinder trajectory
does not pass through the $\Lambda$CDM point. The interacting
trajectories can approach and cross this point at intermediate
redshifts, as highlighted by the inset of Fig.~\ref{fig:sr}. Such
crossings represent instantaneous agreement in the statefinder
plane rather than complete degeneracy with $\Lambda$CDM. The
trajectories consequently retain a geometric signature of the
Kaniadakis deformation and the interaction.
\section{Discussion of the limiting behavior}\label{sec:limits}
The limiting behavior is central to the physical interpretation of
the model. First, the Kaniadakis entropy is even in the
deformation parameter at the order retained here, so the leading
background corrections depend on $K^2$. Second, $K\to0$ implies
$\alpha\to0$ and removes both the positive-power holographic
correction and the entropy-induced Friedmann correction. The
effective density becomes
\begin{equation}
\rho_{\rm DE}\rightarrow3c^2M_p^2R_P^{-2},
\end{equation}
and the particle horizon itself is unchanged. Hence the model
reduces to standard particle-horizon HDE in standard Einstein
gravity. This limit makes the main physical point transparent. The
standard Einstein-gravity particle-horizon HDE model is
nonaccelerating in the absence of interaction. Therefore, the
acceleration obtained here at $K\neq0$ and $b^2=0$ cannot be
attributed to the particle horizon itself. It arises from the
Kaniadakis entropy deformation and the associated modification of
the gravitational background. In this sense the result is a
genuine entropy-induced departure from the standard
particle-horizon HDE behavior. Third, the physical algebraic
branch in Eq.~(\ref{eq:RPbranch}) is selected precisely by the
requirement that it remain finite and reduce to the standard
particle-horizon solution as $K\to0$. The second root is discarded
because it does not possess this physical undeformed limit.
Finally, $b^2\to0$ continuously switches off the dark-sector
interaction. The noninteracting accelerating Kaniadakis solution
is therefore a distinct limit of the full model and should not be
confused with acceleration generated by the interaction.
\section{Conclusions}\label{sec:conclusion}
We have developed a Kaniadakis holographic dark-energy model in a
spatially flat Friedmann-Robertson-Walker universe by choosing the
particle horizon as the infrared cutoff and incorporating the
Kaniadakis correction into both the holographic energy density and
the gravitational background equations. The construction retains
the standard particle-horizon prescription while modifying the
entropy-area relation and the corresponding Friedmann dynamics.

The central physical distinction of this work is between standard
particle-horizon HDE in Einstein gravity and its Kaniadakis
extension. In standard Einstein gravity, when the usual
holographic density $\rho_{\rm HDE}=3c^2M_p^2R_P^{-2}$ is combined
with the particle horizon and the dark sectors are noninteracting,
the equation of state is not sufficiently negative to produce
accelerated expansion. In the dark-energy-dominated limit,
$w_{\rm DE}=-1/3+2/(3c)>-1/3$. Thus, the standard noninteracting
particle-horizon HDE model is decelerating. This provides a clean
reference point for the present analysis. The principal new result
is that the Kaniadakis deformation changes this conclusion without
changing the IR cutoff and without requiring an interaction.

For the numerical analysis we fix
$\Omega_{\rm DE0}=0.7$, $c^2=0.64$, and the representative
Kaniadakis parameter $K=1.90\times10^{-36}$. This value is adopted
as a representative choice that places the deceleration-to-
acceleration transition close to $z\simeq0.6$. With the parameter
set fixed, the noninteracting solution gives
$z_t\simeq0.585$, where $q$ changes sign. The interacting cases
remain remarkably close to this value: $z_t\simeq0.586$ for
$b^2=0.03$ and $z_t\simeq0.589$ for $b^2=0.06$. Therefore, in
contrast to the earlier interpretation in which the interaction was
said to move the transition substantially toward lower redshift,
the updated numerical results show only a very small shift, and it
is toward slightly higher redshift. Most importantly, acceleration
is already present for $b^2=0$; the interaction is not the mechanism
responsible for the onset of acceleration.

The background quantities show that the dark-energy fraction grows
toward the present epoch, while the interaction produces only modest
changes in the evolution of $\Omega_{\rm DE}$ and
$w_{\rm DE}$. The total equation of state remains negative at late
times, consistently with the accelerated phase indicated by the
zero crossing of $q$. The close agreement of the three transition
redshifts also demonstrates that the main qualitative conclusion is
not driven by the phenomenological interaction.

The stability analysis provides an important qualification. The
adiabatic squared sound speed $v_s^2$ becomes negative over
substantial portions of the redshift range displayed in Fig.~\ref{fig:vs2},
with a pronounced singular behavior at an intermediate redshift.
The negative behavior becomes more pronounced as the interaction
strength is increased, especially for $b^2=0.06$. Hence, the
existence of a viable accelerating background should not be
interpreted as evidence of complete stability at the perturbative
level. A full perturbation analysis of the underlying
Kaniadakis-modified gravitational theory is required to determine
the physical significance of this effective-fluid instability.

The statefinder analysis provides an additional geometrical
discriminator. The noninteracting KHDE trajectory does not pass
through the $\Lambda$CDM fixed point $\{r,s\}=\{1,0\}$, while
some interacting trajectories can cross or approach this point at
intermediate redshifts. Such behavior represents instantaneous
agreement in the statefinder plane rather than full dynamical
degeneracy with $\Lambda$CDM. The $s$--$r$ trajectories therefore
retain information about both the Kaniadakis deformation and the
dark-sector interaction.

The $K\to0$ limit provides an essential consistency check. In this
limit, $\alpha\to0$, the positive-power Kaniadakis contribution
disappears, and the effective density reduces continuously to
$3c^2M_p^2R_P^{-2}$. The physical algebraic branch of the
particle-horizon equation is also selected by this continuity
requirement. The undeformed theory is consequently the standard
particle-horizon HDE model in Einstein gravity, including its
nonaccelerating noninteracting behavior. The accelerated solution at
finite $K$ is therefore a genuine consequence of the entropy-induced
modification while the IR cutoff is kept fixed.

In summary, the updated numerical analysis supports the following
main conclusion: \emph{for $K=1.90\times10^{-36}$, the Kaniadakis
correction allows noninteracting particle-horizon HDE to evolve from
a decelerating matter-dominated regime to late-time acceleration,
with $q=0$ at $z\simeq0.585$; the interaction strengths considered
here change this transition only marginally, to $z\simeq0.586$ and
$0.589$.} This result distinguishes the present mechanism from
acceleration generated primarily by dark-sector energy transfer.
At the same time, the sound-speed behavior emphasizes that
background acceleration alone is insufficient to establish the full
physical viability of the model.

The present work is restricted to homogeneous background evolution
and an adiabatic effective-fluid stability diagnostic. A complete
assessment requires a perturbation analysis and a statistical
confrontation with Type-Ia supernovae, BAO, CMB, and expansion-rate
data. Such an analysis would test whether the parameter region that
yields particle-horizon acceleration remains viable against current
observations and whether the Kaniadakis mechanism can provide a
competitive alternative to $\Lambda$CDM and other entropy-deformed
holographic models \cite{ZhangWu2005}.
\section*{Acknowledgements}
We thank Shiraz University Research Council.

\end{document}